\documentclass{aa}
\usepackage{graphicx}
\usepackage{multirow}
\usepackage{float}
\usepackage{txfonts}
\usepackage{hyperref}
\usepackage[normalem]{ulem}
\usepackage{natbib}
\bibpunct{(}{)}{;}{a}{}{,}	
\begin{document}

\title{The intrinsic Baldwin effect in broad Balmer lines of six long-term monitored AGNs}
\titlerunning {Intrinsic Baldwin effect in six AGNs}
\author{N. Raki\'c\inst{1,2}, G. La Mura \inst{3}, D. Ili\'c\inst{2,4},  A.I. Shapovalova\inst{5}, W. Kollatschny\inst{6}, P. Rafanelli \inst{3}, L. {\v C}. Popovi\'c\inst{1,2,7}}
\institute{Faculty of Science, University of Banjaluka, Mladena Stojanovi\'ca 2,78000 Banjaluka, Republic of Srpska, Bosnia and Herzegovina; \email{nemanja.rakic@unibl.rs}
\and
Department of Astronomy, Faculty of Mathematics, University of Belgrade, Studentski Trg 16, 11000 Belgrade, Serbia
\and 
Department of Physics and Astronomy, University of Padova, vicolo dell'Osservatorio 3, I-35122 Padova, Italy
\and Isaac Newton Institute of Chile, Yugoslavia Branch
\and 
Special Astrophysical Observatory of the Russian Academy of Science,
Nizhnij Arkhyz, Karachaevo-Cherkesia 369167, Russia
\and 
Institut fur Astrophysik, Universitat Gottingen, Friedrich-Hund Platz 1, 37077, G\"{o}ttingen, Germany
\and
Astronomical Observatory, Volgina 7, 11060 Belgrade, Serbia}
\date{}
\authorrunning{Raki\'c et al.}

\abstract{}

\abstract {We investigate the intrinsic Baldwin effect (Beff) of the broad H$\alpha$ and H$\beta$ emission lines 
for six Type 1 active galactic nuclei (AGNs) with different broad line characteristics: two Seyfert 1 (\object{NGC 4151} and \object{NGC 5548}), two AGNs with double-peaked broad line profiles (\object{3C 390.3} and \object{Arp 102B}), one narrow line Seyfert 1 (\object{Ark 564}), and one high-luminosity quasar with highly red asymmetric broad line profiles (\object{E1821+643}). We find that a significant intrinsic Beff was present in all Type 1 AGNs in our sample. Moreover, we do not see strong differences in intrinsic Beff slopes in different types of AGNs, which probably have different physical properties such as inclination, broad line region (BLR) geometry, or accretion rate. Additionally, we find that the intrinsic Beff was not connected with the global  Beff, which, instead, could not be detected in the broad H$\alpha$ or H$\beta$ emission lines.  In the case of \object{NGC 4151}, the detected variation of the Beff slope could be due to the change in the site of line formation in the BLR.  Finally, the intrinsic Beff
might be caused by the additional optical continuum component that is not part of the ionization continuum. }


\maketitle
\section{Introduction}

Correlations between the continuum and line emission of active galactic nuclei (AGNs) are a valuable probe of the physics and structure of the central engine and its close environment. For a sample of objects, \citet{ba77} found that the equivalent width (EW) of the C IV~$\lambda 1549$ emission line was anti-correlated with the luminosity of the underlying continuum. This property, named ``the Baldwin effect (Beff)'' by \citet{cs78}, was detected in almost all UV and optical broad emission lines  \citep[see, e.g.,][]{di02,do09,kov10,sl15}, in the narrow lines \citep[see][for a review]{sh07} though with different strengths, and also in the X-ray spectra of AGNs \citep[see, e.g.,][]{ji06,ji07,shu12,ri13}.

Today, two different types of Baldwin effect are present in the literature:
        \begin{enumerate}

 \item {\it global or ensemble Baldwin effect} -- anti-correlation between the EW of the emission line and  the underlying continuum luminosity of single-epoch observations of a large number of AGNs. 
 \item {\it intrinsic Baldwin effect} -- anti-correlation between the EW of the emission line and the underlying continuum of individual variable AGNs \citep{pp92}.

\end{enumerate}

The existence of an intrinsic correlation and its relation with the global effect provides important clues about how the broad emission lines are produced, and it is therefore
important in order to better constrain the nature of the broad line region (BLR).

For this reason, the Beff has been investigated in several papers and different mechanisms have been 
proposed as possible interpretations. In spite of the numerous efforts to clarify its nature, 
the question of the physical meaning of the Beff has long been a matter of debate 
\citep[see, e.g.,][]{bg92,ne92,zh92,zm93,gr96,gr98,ko98,wa03,ba04,bl04,xu08,pk11,bi12}.
 
The main  explanations that  have been suggested are as follows:
\begin{description}
\item{(i)} a luminosity dependence of the ionization parameter $U$ and an anti-correlation between the continuum luminosity and the BLR covering factor \citep{mf84};
\item{(ii)} a geometrical effect due to the combination of an inclination dependent anisotropic continuum with a more isotropic line emission \citep{ne85};
\item{(iii)} a different variability pattern in the thermal and nonthermal components of the continuum, leading to changes in the spectral energy distribution (SED) of the ionizing photons, which affect the broad line intensities independently of the state of the continuum close to the line
\citep{ki90};
\item{(iv)} a multiple component nature of the broad emission lines, with part of the line flux not directly
controlled by the continuum luminosity \citep[][]{pp92,sh08};
\item{(v)} a general trend of softer continuum shapes and enhanced metallicities at higher
luminosities \citep{di02};
\item{(vi)} a different Eddington ratio \citep[see][]{ca04,bl04,ba04,ma08,do09,bi12} or mass of the supermassive black hole (SMBH) \citet{xu08}. 
\end{description}
Different combinations of the fundamental effects listed above can be used to explain both the intrinsic and global Beff.

 In the case of the global Beff, it has been established that the higher ionization lines  show steeper slopes in the EW versus $L$ plot \citep[see Fig. 9 in][]{di02}. Some investigations have revealed that the Balmer lines show no global Beff \citep[see, e.g.,][]{kov10,pk11} or even a weak inverse global Beff \citep[positive correlation between a line and corresponding continuum][]{cro02,gr05}. \citet{za92} combined geometrical effects with the luminosity dependent ionization parameter $U$ to explain the Beff of high- and low-ionization lines, whereas \citet{zm93} suggested that the most luminous objects could produce relatively softer continua, resulting in weaker high-ionization lines. In general, the possibility that different SMBH masses and accretion rates, together with changes in the gas metallicity, could be combined to provide an interpretation of the global Beff in the framework of cosmological evolution is often suggested \citep[][]{ko99,gr01,sh03}.

On the other hand, the intrinsic Beff has been studied in several papers, but only for strongly variable AGNs. \citet{ki90} observed  a set of seven objects observed with the IUE satellite for more than 15 epochs, and found the presence of the intrinsic Beff in the UV line C IV and Ly$\alpha$. \citet{gp03} showed that the intrinsic Beff is present in the broad H$\beta$ line of NGC~5548. They considered the light travel times from the continuum source to the line emitting region in NGC~5548 and found that an intrinsic Beff held even when comparing the broad line EWs with the continuum that actually affected the gas at the time of the line emission. \citet{go04} showed that photoionization models predicted a time variable response of emission lines  from an extended BLR that was entirely consistent with H$\beta$ line observations.
Interestingly,  variations in the slope of the intrinsic Beff have been found so far for the C IV line in \object{NGC 4151}  \citep{ko06} and Fairall 9 \citep{wa86,os99}, and for H$\beta$ in \object{NGC 5548} \citep{go04}. This shows that a nonconstant slope may be common for AGNs. Moreover, the intrinsic Beff has a steeper slope than global Beff \citep{ki90,pp92}. An additional detail seen in the intrinsic Beff is that the slope in log(EW) versus $\log L$ relation steepens when the source is becoming more luminous \citep[see,  e.g.,][]{ko06}. 
This trend of the slope steepening for higher continuum luminosities is also observed in the global Beff
\citep{os99, bl04}. However, the global and intrinsic Baldwin effects probably have completely different physical origins, thus they are unlikely to be connected. For the intrinsic Beff, several authors have advocated the idea that it could be explained by the different phases in the activity of an accretion powered source \citep{wa99b,wa99a,ko06}.

The aim of this work is to investigate the intrinsic broad line Beff in Type 1 AGNs with different characteristics in order to explore the effect of the BLR orientation and geometry on the intrinsic Beff and to discuss the lack of global Beff in the broad Balmer lines. Therefore,  we combine observations of six long-term monitored objects: two typical Seyfert 1 galaxies (\object{NGC 4151} and \object{NGC5548}), two AGNs with double-peaked broad lines (\object{3C 390.3} and Arp~102B), one narrow line Seyfert 1 -- NLSy1 galaxy  (\object{Ark 564}), and one high-luminosity quasar with highly red asymmetric broad line profiles (\object{E1821+643}). We explore the intrinsic Beff in the broad H$\alpha$ and H$\beta$ emission lines, and the H$\alpha$ and H$\beta$ ratio as a function of the continuum flux.  We discuss the intrinsic Beff together with the global  Beff. Finally, we speculate on the physical background of the intrinsic Beff.

The paper is organized as follows. In Sect. 2 we describe the sources of used data and the method of analysis. 
Our results and the discussion are presented in \S 3, and  in \S 4 we discuss some possible physical processes that cause the intrinsic Beff. Finally, in \S 5 we outline our conclusions.

\section{Data and method of analysis}
\label{sec:data}

To construct a uniform and consistent  data set for all AGNs, we used the observational data with already measured line and continuum fluxes from our monitoring campaigns \citep[see][and references therein for more details]{il15}. The sample contains six long-term monitored AGNs with different spectral characteristics and subclassifications:
\begin{enumerate}
\item  Seyfert 1 AGNs  \object{NGC 5548} and \object{NGC 4151}, which showed strong variability in the broad spectral lines (full width at half maximum (FWHM) $\sim 5000-7000 \rm \ km s^{-1}$) and continuum flux \citep[see, e.g.,][]{sh04,sh08,sh10a}. Moreover, their broad line profiles exhibit noticeable change in  line shape and intensity, resulting in periods when these AGNs almost show the characteristics of Seyfert 2 
\citep[very weak broad lines; see][]{sh04,sh08,sh10a};
\item Double-peaked broad line AGNs  \object{3C 390.3} and \object{Arp 102B},  whose broad Balmer lines show two distinguished peaks with very large overall width (FWHM $\sim 10000 \rm \ kms^{-1}$). This type of objects are called double-peaked line (DPL) AGNs.  These lines are supposed to be emitted from an accretion disk \citep{ch89}. \object{3C 390.3}, which is a radio loud AGN, seems to have higher variability in the spectra \citep[][]{sh10b} than \object{Arp 102B} \citep[][]{sh13}; 
\item Narrow line Seyfert 1 (NLSy1) AGN  \object{Ark 564}, which has a strong Fe II emission and 
somewhat narrower broad lines (FWHM $\sim 2000 \rm\ kms^{-1}$) with a prominent narrow component. The variability of this AGN is weak, but  there are occasional flare-like outbursts \citep[see][]{sh12};
\item High-luminosity quasar  \object{E1821+643}, a low-redshift ($z$=0.297) radio-quiet quasar. Its broad emission lines (FWHM $> 5000$ $\rm \ kms^{-1}$) are redshifted by $\sim 1000$ km s$^{-1}$ and have strongly asymmetric profiles in the red part. These effects make this object a candidate for either a supermassive binary black hole (SMBBH) or a recoiling black hole after SMBH collision \citep{ro10, sh16}. It is a weakly variable quasar, but  some flare-like outbursts have been detected \citep[see][]{sh16}.
\end{enumerate}

For the comparison of the intrinsic Beff in our sample with the global Beff trend, we used the spectra of 21416 Type 1 AGNs taken from the Sloan Digital Sky Survey (SDSS) database  studied by \cite{shen11}, combined with a sample of 90 low-luminosity objects, also from the SDSS database, analyzed in detail by \cite{la07} in order to cover a wider luminosity range. Furthermore, from \cite{shen11} we extracted the subsample of 4800 broad line AGNs to study the H$\alpha$/H$\beta$ ratio. 

We adopt the cosmological parameters used in \cite{shen11}, 
$\rm H_0=70$ $\rm km\,s^{-1}\,Mpc$ , $\rm \Omega_{matter}=0.30$, and $\Omega_{\rm vacuum}=0.70$ when calculating the luminosity and luminosity distance.

\subsection{Data reduction}
\begin{table*}
\caption{Basic information about the observations of the used spectra.\label{obs}}
\tiny
\centering
\begin{tabular}{ccccccccc}
\hline\hline
Object$^{(a)}$ & Type$^{(b)}$ & Period$^{(c)}$ &Telescope$^{(d)}$ & Aperture$^{(e)}$ & Line$^{(f)}$& Continuum$^{(g)}$& No.$^{(h)}$ & Reference$^{(i)}$\\\hline

\object{NGC 5548} & Seyfert 1 & 1996--2002 & SAO, GHAO, OAN-SPM & $4.2" \times19.8"$ & H$\alpha$ & 5100 \AA  & 23 & 1 \\
             &     &&        &                       & H$\beta$  & 5100 \AA  & 81 &  \\    
\object{NGC 4151} & Seyfert 1 & 1996--2006 & SAO, GHAO, OAN-SPM &$2.0"\times6.0"$& H$\alpha$ & 5100 \AA  & 137 & 2 \\
                 &   & &          &                   & H$\beta$  & 5100 \AA  & 181 & \\         
\object{3C 390.3} & DPL &1995--2007 & SAO, GHAO, OAN-SPM& $2.0" \times 6.0"$& H$\beta$ & 5100 \AA & 128& 3 \\ 
\object{Arp 102B} & DPL &1987--2013 & SAO, GHAO, OAN-SPM, CA & $2.5" \times 6.0"$& H$\alpha$ & 6200 \AA & 88 & 4 \\
                  &                         &&&& H$\beta$ & 5100 \AA & 116&  \\
\object{Ark 564}  & NLSy1 & 1999--2010 & SAO, GHAO, OAN-SPM& $2.5" \times 6.0"$& H$\alpha$ & 6200 \AA & 50& 5 \\
            &&       &                       &   & H$\beta$ & 5100 \AA & 91& \\
\object{E1821+643} &quasar&1990--2014& SAO, GHAO, CA& $2" \times 6.0"$& H$\beta$&  5100 \AA & 127& 6 \\\hline
\end{tabular}
\tablefoot{Columns: (a) Object name; (b) AGN type; (c) Monitored period; 
(d)  Telescope used (SAO - 6m and 1m telescopes of Special Astrophysical Observatory; 
GHAO - 2.1m telescope of Guillermo Haro Astrophysical Observatory; OAN-SPM - 2.1m telescope of the 
Observatorio Astron\'{o}mico Nacional at San Pedro Martir; and CA - 3.5m and 2.2m telescopes of Calar Alto); (e)  Projected spectrograph entrance apertures (slit width$\times$slit length in
arcsec); (f)  Emission line; (g)  Continuum; (h)  Number of spectra; (i)   Reference from which line and continuum fluxes are taken. \\
\tablebib{
(1)~\citet{sh04};
(2) \citet{sh08};
(3) \citet{sh10b};
(4) \citet{sh13};
(5) \citet{sh12};
(6) \citet{sh16}.}
}

\end{table*}

\begin{figure}
        \centering
        \includegraphics[width=\columnwidth]{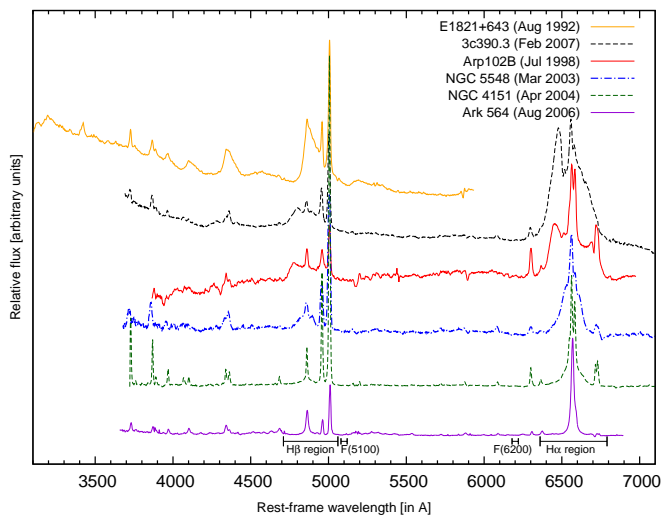}
        \caption{Spectra of six studied objects (as listed in top right
corner, object name and observation epoch are given). Intensity is given in arbitrary units shown for comparison. The measured continuum and line extraction bands are marked below the first  spectrum.}
        \label{spectrum}
\end{figure} 

Optical spectral data used in the investigation of the intrinsic Beff are taken from the published long-term monitoring campaigns described in several papers \citep[see][]{sh04,sh08,sh10b,sh12,sh13,sh16}, where a detailed description of the data reduction  and flux measurements was given and therefore is not repeated here. 
 We emphasize  that all spectral data were reduced and analyzed using the same method in order to construct a uniform data set in a consistent way. We list below only some basic information.
The spectra were obtained with the following instruments:
(i) the 6 m and 1 m telescopes of the Special Astrophysical Observatory (SAO), Russia; (ii) the 2.1 m telescope of the 
Guillermo Haro Astrophysical Observatory (GHAO) at Cananea, Sonora, Mexico; (iii) the 2.1 m telescope of 
the Observatorio Astron\'{o}mico Nacional at San Pedro Martir (OAN-SPM), Baja California, Mexico; and 
(iv) the 3.5 m and 2.2 m telescopes of the Calar Alto Observatory (CA), Almer\'{i}a Spain. The typical properties of the observed spectra 
were the wavelength range 4000 - 7500 $\mbox{\AA}$, the spectral resolution 
$\Delta\lambda=5-15 \mbox{\AA}$, and the S/N ratio $>$ 50 in the continuum near $\rm H\alpha$ and $\rm H\beta$.
All spectra, taken with different telescopes, were flux-calibrated by scaling the overall spectra to have the same AGN narrow emission line fluxes. The typical line and continuum flux uncertainties are $\sim$3-5\%. The narrow line flux does not vary because the  narrow line region (NLR) is ten to kpc scale, thus its variation in years is negligible \citep[see, e.g.,][]{sh04,sh08,sh10b,sh12,sh13,sh16,runn16}. However, different  aperture sizes will result in different narrow line fluxes, and consequently  can affect the scaled continuum and broad line fluxes. Therefore,  special attention is given to the correction of the continuum and broad line flux to aperture effects \citep[see][]{pe95,pe02,sh01,se02,sh13,sh16}, and we used the observed data that have been corrected for this effect. A summary of the observations and  the data used is listed in Table \ref{obs}. For  illustration, in Fig. \ref{spectrum} we plot an example spectrum of each studied object in arbitrary units for comparison, shown in the order  listed in the top right corner of Fig. \ref{spectrum}.

\subsubsection{Host-galaxy and narrow-line contribution}

For the studies of the intrinsic Beff, it is important to correct the  continuum fluxes for the host-galaxy contribution and line fluxes for the narrow-line contribution.

 For the  host-galaxy contribution to the continuum flux at 5100 \AA\, and 6200 \AA , 
we took the estimates for a given aperture size from the literature (see references in Table \ref{tab:galaxy} for details of the procedures used). The estimated host-galaxy flux, the aperture size, and the references are listed in Table \ref{tab:galaxy}. For most objects, the aperture size used for our flux measurements is similar to the aperture size used to estimate the host-galaxy flux (see Tables \ref{obs} and \ref{tab:galaxy}), thus we simply subtracted the host-galaxy flux from the continuum flux. Only in the case of \object{NGC 5548} did we first correct the continuum flux to the aperture of $5"\times 7.5"$ used by \citet{pe95, pe02} for the host-galaxy contribution. For the correction of the continuum at 6200 \AA\, of \object{Ark 564}, we calculated the scaled value from the host-galaxy flux at 5100 \AA\  as that was only available in the literature \citep{sh01}.

The narrow-line contribution to the $\rm H\alpha$ and $\rm H\beta$ emission 
lines was already subtracted in the published data for \object{NGC 4151}, \object{NGC 5548}, \object{3C 390.3}, and \object{E1821+643} \citep{sh04, sh08, sh10b,sh16}. For \object{Arp 102B}, we took the \citet{sh13} estimated values:  $5.46\times10^{-14}$ $\rm erg\,cm^{-2}\,s^{-1}$  for the narrow H$\beta$ + [O III]$\lambda\lambda$4959,5007 lines, and $13.52\times10^{-14}$ $\rm erg\,cm^{-2}\,s^{-1}$ for the    
narrow H$\alpha$+[NII]$\lambda\lambda$6548,6584 + [S II]$\lambda\lambda$6717,6731 lines. For the case of \object{Ark 564} we took the \citet{sh12} estimates:  20\% narrow-line contribution to the mean H$\rm \beta$ flux, and 30\% to the mean H$\rm \alpha$ flux.

\begin{table*}

\caption{Properties of the AGNs in our sample used in the paper.\label{tab:galaxy}}
\tiny
\centering
\begin{tabular}{cccccccccc}
\hline\hline
Object$^{(a)}$ & Aperture$^{(b)}$ & $F_{\rm gal}^{(c)}$ & Reference$^{(d)}$ & FWHM$^{(e)}$ & 
$R_{\rm BLR}^{(f)}$ & $F_{\rm max} / F_{\rm min}^{(g)}$  & Reference$^{(h)}$ & JD(div)$^{(i)}$ & 
$F_{\rm cnt}$(div)$^{(j)}$ \\
 & & blue/red & &H$\rm \beta$/H$\rm \alpha$ & H$\rm \beta$/H$\rm \alpha$ & H$\rm \beta$/H$\rm \alpha$ 
 & & 2400000+ &\\
\hline
\object{NGC 4151} & $2.0"\times6.0"$ &$6.00$& 1  &6110/4650 &5/6 & 4.8/3.1 & 1 & 51283 & 46.2  \\
\object{NGC 5548} & $5.0"\times7.5"$ &$3.06$& 2,3  & 6300   & 49/27 & 4.9/3.5  & 8 & 51702  &4.41 \\
\object{3C 390.3} & $2.0"\times6.0"$ &$0.46$ & 4 & 11900/11000 & 96/120 & 4.7/3.4 & 9 & 52327 & 0.74 \\
\object{Arp 102B} & $2.5"\times6.0"$ &$1.13$/$1.21$ & 5 & 15900/14300 & 15/21 & 3.0/2.4 & 5 &55367  & 0.38 \\
\object{Ark 564}  & $2.5"\times6.0"$ & $2.40/0.99$ & 6 & 960/800& 4/5& 1.6/1.5  & 10 & 53505 &2.55  \\
\object{E1821+643} &$2.0"\times6.0"$ &0.29& 7& 5610& 120& 1.4& 7 & 53683 &10.35  \\
\hline
\end{tabular}
\tablefoot{Columns: (a)  Object name; (b)  Projected spectrograph entrance aperture for the host galaxy; (c)  Host-galaxy contribution in units of $10^{-15}\rm erg\,cm^{-2}\,s^{-1}\, \mbox{\AA}^{-1}$; (d)  References from which the entrance aperture and $\rm F_{galaxy}$ are taken; (e)  FWHM of the mean line profile in units of km s$^{-1}$; (f)  Photometric BLR radius in light days; (g)   Maximum to minimum line flux ratio; (h)  References from which the FWHM, $\rm R_{BLR}$, and $F_{\rm max} / F_{\rm min}$ are taken; (i-j)  Modified Julian date dividing  period 1 and 2 of different levels of activity and the
corresponding continuum flux at 5100 \AA\ in units of $10^{-15}\rm erg\,cm^{-2}\,s^{-1}\, \mbox{\AA}^{-1}$.}\\
\tablebib{(1)~\citet{sh08};
(2,3) \citet{pe95,pe02};
(4) \citet{se02};
(5) \citet{sh13};
(6) \citet{sh01};
(7) \citet{sh16};
(8) \citet{sh04};
(9) \citet{sh10a};
(10) \citet{sh12}.}

\end{table*}

\subsection{Method of analysis}
\label{sec:meth}

The  Beff can be expressed as a simple power law relation:
\begin{equation}
        EW_{\lambda} \propto L_{\lambda}^{\beta},
        \label{eq:1}
\end{equation}
where $EW_{\lambda}$ is the equivalent width of the emission line (in $\mbox{\AA}$), $L_{\lambda}$ is 
the corresponding specific continuum luminosity (in $\rm erg$ $\rm s^{-1}$ $\mbox{\AA}^{-1}$) at 
wavelength close to the emission line, and $\beta$ is the slope. Hence, in order to have a Beff, the value must be $\beta<0$.

As we investigate the intrinsic Beff of a specific object, the luminosity can be replaced with the flux.  It is very important to calculate EW in the way that represents the response of an emission line to the continuum that actually stroke the BLR at the line emission time \citep[see, e.g.,][]{gp03,gk14}. Thus, we took into account the time delays due to light travel time from the continuum source to the BLR for each object obtained from the reverberation mapping. For this we first interpolated the continuum and emission line light curves, and then calculated the EW using the line flux that is corrected for the time delay (listed in Table \ref{obs} as photometric BLR radius) with respect to the continuum flux.
We define the equivalent width of the $\rm H\beta$ and $\rm H\alpha$ lines as follows:
\begin{equation}
        \rm EW=\frac{F_{line}({\rm JD}+\tau)}{F_{cnt}({\rm JD})}
        \label{eq:ew}
,\end{equation}
where $F_{\rm cnt}({\rm JD})$ represents the continuum flux for a certain epoch and $F_{\rm line} ({\rm JD} + \tau)$ is the broad line flux at the epoch corrected for the time delay $\tau$.
In those cases where the continuum close to the H$\alpha$ line was not measured, we used the blue continuum close to the H$\beta$ line (see Table \ref{obs}). 

Assuming that the data follow a relationship of the form
\begin{equation}
        \rm F_{line} = k F_{cnt}^{\alpha} \,,
\end{equation}
and taking into account Eq. (\ref{eq:ew}), we performed the  least-squares fits to the linear function:
\begin{equation}
        \rm \log EW_{\lambda}=A + \beta \log(F_{cnt}),
        \label{eq:fit}
\end{equation} 
where $A=\log k$ is a normalization constant and $\beta=\alpha-1$ is the best-fitting slope of the intrinsic Beff. 
We calculated the Pearson correlation coefficients $r$ for each fit that indicated the rate of correlation and its statistical significance.

{ Furthermore, to be able to compare and discuss different objects,  in Table 
\ref{tab:galaxy} we list parameters for each galaxy such as the line FWHM, the reverberation photometric radius of the BLR ($R_{\rm BLR}$), and the ratio of the maximum to minimum line flux for H$\beta$ and H$\alpha$. The last parameter is an indication of the object variability.}

\section{Results and discussion}

Here we present the intrinsic Beff, shown as  $\log$(EW) versus $\log(F_{\rm cnt}),$ for the broad Balmer lines of six AGNs. Additionally, we explore how the H$\alpha$/H$\beta$ flux ratio changes as a function of the 
continuum in order to discuss physical background of the intrinsic Beff. We conclude this section with a comparison of the intrinsic Beff  of the six monitored AGNs with the global Beff.

\subsection{Intrinsic Beff}
\begin{table*}
\caption{Parameters of the intrinsic Beff.\label{res}}
\tiny
\centering
\begin{tabular}{lcccccc}
\hline\hline
Object$^{(a)}$ & Line$^{(b)}$ & $r$$^{(c)}$ & $P$$^{(d)}$ & $\rm \beta$$^{(e)}$ & $A$$^{(f)}$ & Data sets$^{(g)}$\\\hline

\multirow{4}{*}{\object{NGC 5548}}&      \multirow{3}{*}{$\rm H\alpha$} & -0.997 & $<0.0001$ & -0.701& -7.31 & p1 \\ 
        && -0.984 & $<0.0001$ & -1.031 & -12.21 & p2\\
        && -0.927 & $<0.0001$ & -0.577 & -5.52 & p1+ p2\\ \cline{2-7}

        &\multirow{2}{*}{$\rm H\beta$} &-0.959 & $<0.0001$ & -0.597 & -6.36& p1\\ 
        && -0.913 & $0.0006$ & -1.284 & -16.47& p2\\
        && -0.936 & $<0.0001$ & -0.568 & -5.96& p1+p2\\

\hline
\multirow{6}{*}{\object{NGC 4151}}&\multirow{3}{*}{$\rm H\alpha$}
& -0.983 & $<0.0001$ & -0.922 & -9.45& p1\\
&&-0.954 & $<0.0001$ & -0.691 & -6.51 & p2\\
&&  -0.953 & $<0.0001$ & -0.580 & -4.98&p1+ p2\\ \cline{2-7}
&\multirow{3}{*}{$\rm H\beta$}& -0.912& $<0.0001$ & -0.617 & -6.05& p1\\
&& -0.886 & $<0.0001$ & -0.513 & -4.78& p2\\
&& -0.892 & $<0.0001$ & -0.392 & -3.11 &p1+ p2\\
\hline

\multirow{6}{*}{\object{3C 390.3}}
&\multirow{3}{*}{$\rm H\alpha$} & -0.715 & 0.0006& -0.452 & -3.90& p1\\
&& -0.793 & $<0.0001$ & -0.394 & -3.07& p2 \\
&& -0.831 & $<0.0001$ & -0.445 & 3.81 &p1+ p2\\\cline{2-7}

&\multirow{3}{*}{$\rm H\beta$} & -0.593 & $<0.0001$& -0.384 & -3.52& p1\\
&& -0.804 & $<0.0001$ & -0.334 & -2.82& p2 \\ 
&& -0.834 & $<0.0001$ & -0.407 & 3.87 &p1+ p2\\
\hline

\multirow{4}{*}{\object{Arp 102B}}&$\rm H\alpha$ & -0.868 & $<0.0001$& -0.912 & -11.17 & p1+p2 \\\cline{2-7}

        &\multirow{3}{*}{$\rm H\beta$} & -0.825& $<0.0001$ & -0.858& -11.02& p1 \\
        
        && -0.956 & 0.0028  & -1.310 & -18.25 & p2\\
        && -0.755 & $<0.0001$& -0.774 & -9.73 & p1+ p2\\ 
        
\hline
        \multirow{6}{*}{\object{Ark 564}}&\multirow{3}{*}{$\rm H\alpha$} 
    &-0.651 & 0.0002 & -0.523 & -5.02 & p1\\
    &&-0.820 & $<0.0001$ & -0.807 & -9.28 & p2\\
    &&--0.884 &  $<0.0001$ & -0.680 & -7.37& p1+p2\\\cline{2-7}         
        &\multirow{3}{*}{$\rm H\beta$} 
        & -0.591&$<0.0001$  & -0.701 & -8.939 & p1\\
        && -0.485& $0.0005$& -0.747 & -9.521& p2\\ 
        && -0.701& $<0.0001$ & -0.577& -7.420& p1+p2 \\\hline
        \multirow{3}{*}{\object{E1821+643}}&\multirow{3}{*}{$\rm H\beta$}
& -0.911 & $<0.0001$ & -0.807 & -9.38& p1\\
&&-0.777 & $<0.0001$ & -0.630 & -6.86 & p2\\
&&-0.879 & $<0.0001$ & -0.658 & -7.25&p1+ p2\\ \hline
\end{tabular}

\tablefoot{Columns: (a) Object; (b) Emission line; (c) Pearson correlation coefficient $r$; 
(d): Null hypothesis probability $P$; (e) Slope of the least-squares linear fit  $\beta$; (f) Constant of the linear-fit $\log A$; (g) Data set denoted with p1 + p2 represents full data range, p1 only period 1, and p2 only period 2.\\}
\end{table*}
\begin{figure*}
        \centering
        \includegraphics[width=0.9\textwidth]{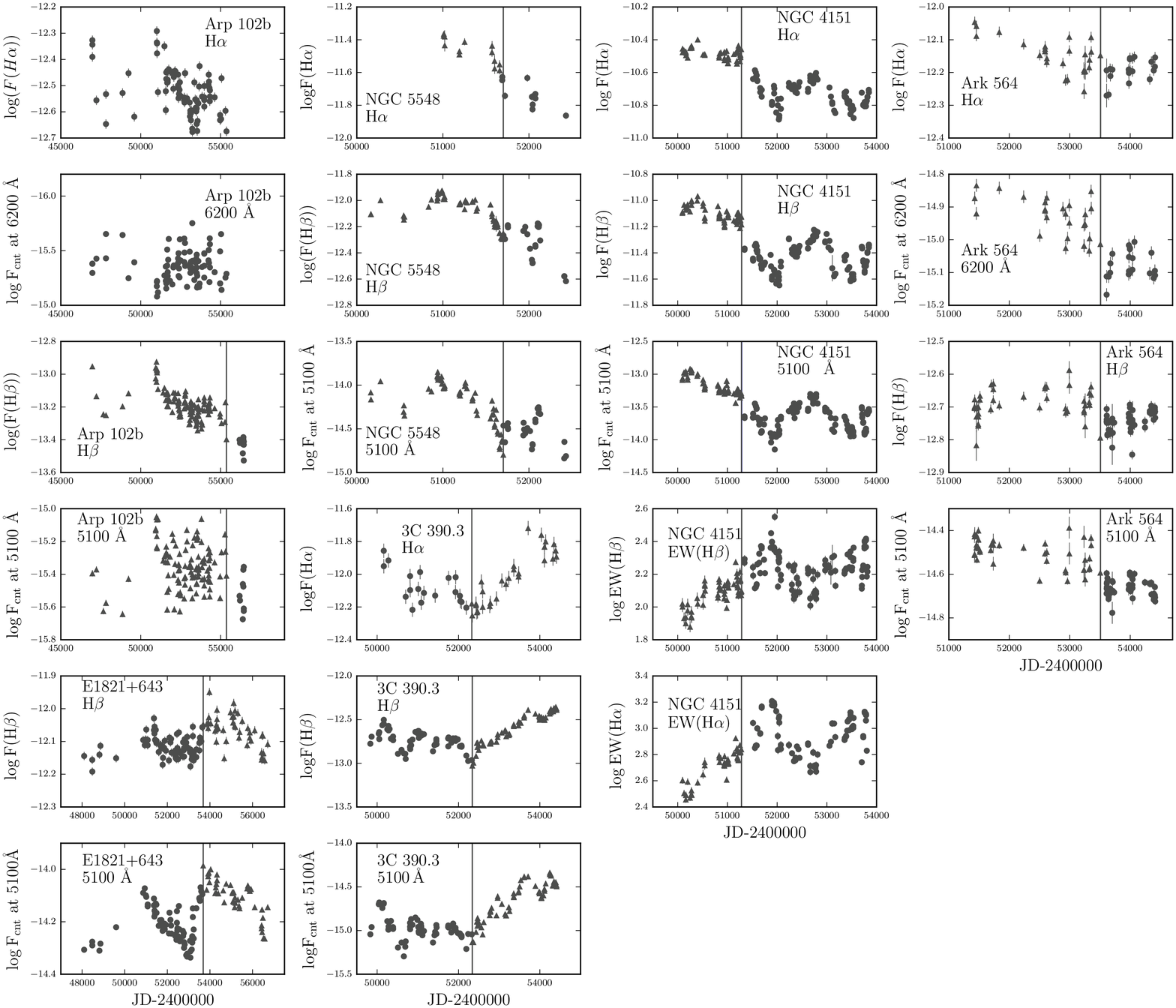}
        \caption{Light curves of the broad H$\alpha$ and H$\beta$ emission line fluxes, together with the continuum fluxes at 5100 $\mbox{\AA}$ and 6200 $\mbox{\AA}$ in logarithmic scale for the sample of six AGNs (object name and line/continuum denoted on each plot). Data are divided on the basis of the high-flux (triangles) and low-flux (full circles)  states. Error bars are also shown, unless smaller than the marker.}
        \label{all}
\end{figure*} 

\begin{figure*}
        \centering
        \includegraphics[width=0.9\textwidth]{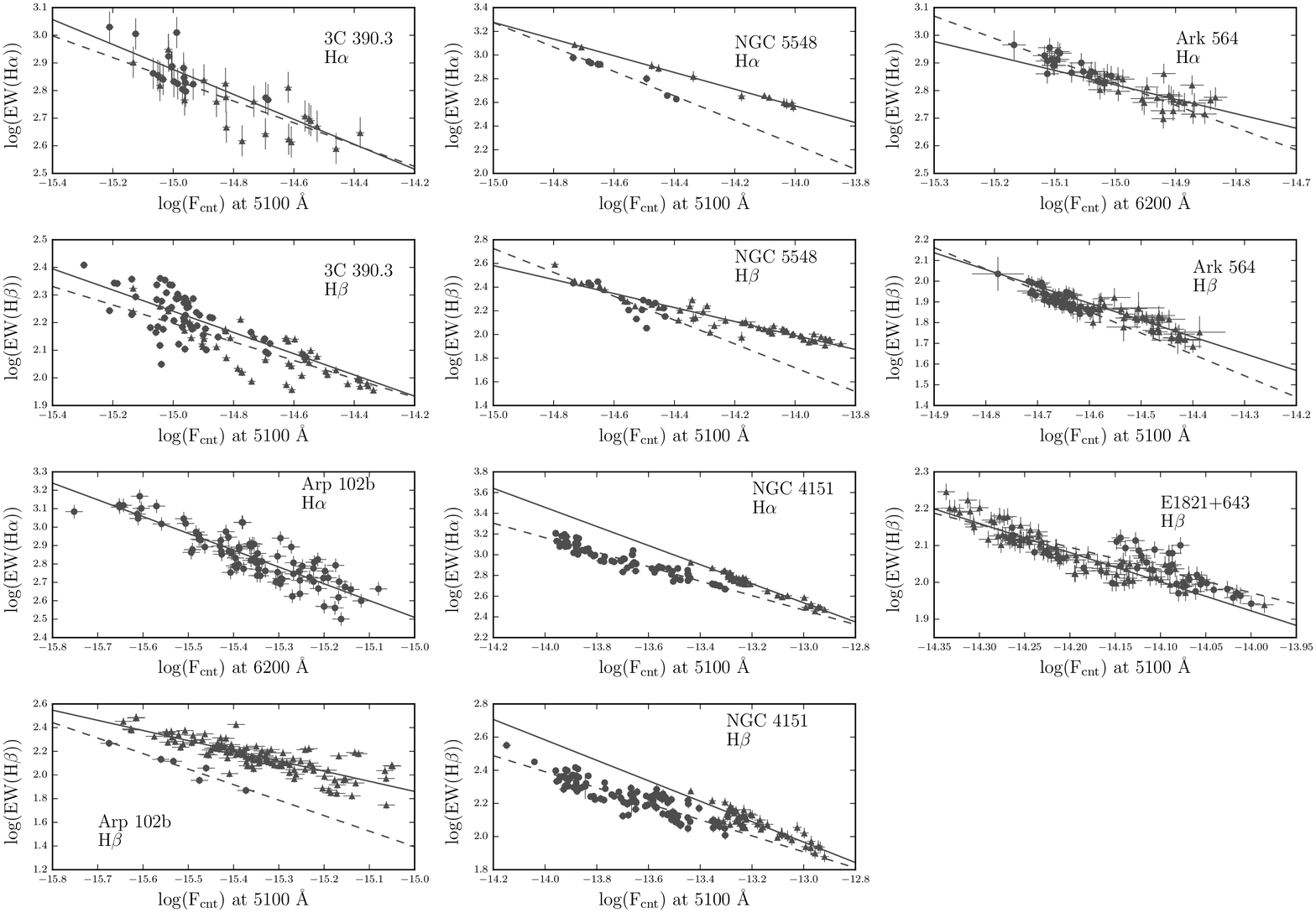}
        \caption{Intrinsic Beff for the sample of six AGNs (name and line denoted on each plot). Data are divided and marked in the same way as in Fig. \ref{all}. Solid and dashed lines represent the best fits to the data. }
        \label{baldwin}
\end{figure*}

The available line and continuum light curves together with the corresponding intrinsic Beff, for our sample of six galaxies, are shown in Figs. \ref{all} and \ref{baldwin}.

In order to calculate the EW and investigate the intrinsic Beff, we used fluxes of the H$\alpha$ and H$\beta$ broad lines and the underlying continuum at 5100 \AA\ (and at 6200 \AA\ where it is available for the H$\alpha$ line). For each object we divided the data set into two subsamples,  which we denote as periods 1 and 2. The periods are defined by the date that distinguishes between the low and high state of activity of the continuum flux at 5100 \AA. The dividing JD and the corresponding continuum flux at 5100 \AA\, are listed in Table \ref{tab:galaxy}.  For all objects we analyze both the overall data set and the separate periods, except for the H$\alpha$ line of \object{Arp 102B}, for which we analyzed only the overall data.
The results of the linear least-squares fit (correlation coefficients $r$ and corresponding $P$-value, slope $\beta$, and normalization constant $A$) between the EW and continuum flux for the broad H$\beta$ and H$\alpha$ lines are given in Table \ref{res}.

Looking into the results presented in Table \ref{res} and Figs. \ref{all} and \ref{baldwin}, we can single out several facts:

(i) {\it Seyfert 1 galaxies  \object{NGC 4151} and \object{NGC 5548}:} There is a strong correlation between $\log$(EW) and the continuum at 5100 \AA\ for both objects. We note that the long-term monitoring of these AGNs show that they have a high degree of variability (with $F_{\rm max}/F_{\rm min}\sim$ 3.5--5 in the line fluxes) and that the broad line profiles are also changing, showing shoulders and two peaks \citep[][]{sh04,sh08,bo12,li16}. Moreover, it was found that both of these AGNs had periodical changes in the continuum and line fluxes which indicated an orbital motion in the BLR, and consequently they were marked as SMBBH candidates  \citep[][]{bo12,popov12,li16,bo16}. The intrinsic Beff slope $\beta$ in these two AGNs is significant for both broad lines in all three analyzed periods, and is always $\beta<-0.4$.

(ii) \textit{Double-peaked emitters  \object{3C 390.3} and \object{Arp 102B}:} They show a significant intrinsic Beff, but the correlation is weaker in \object{3C 390.3} than in \object{Arp 102B}. In addition, the correlations are weaker than in the two  Seyfert 1 AGNs. There is a difference in the slope:  \object{3C 390.3} shows a flatter slope ($\beta \sim-0.4$) than \object{Arp 102B} ($\beta \sim-0.9$). We note that the long-term monitoring of these two galaxies reveals different spectral properties that can be explained as a different structure of the BLR.  \object{3C 390.3} shows a higher rate of variability than \object{Arp 102B} (Table \ref{tab:galaxy}) and
has a typical line variability that can be explained with the disk emission \citep[see][]{sh10a,jo10,po11} and perturbations in the disk, whereas this is not seen in \object{Arp 102B} \citep[see][]{sh13,po14}. In the case of \object{Arp 102B}, the data dispersion  in the Beff plot is much greater. It could be that the host-galaxy contribution to the continuum emission was larger than estimated, leading to a steeper slope of the intrinsic Beff (see Section 4.2). Finally, it is well known that \object{3C 390.3} contains  various radio structures \citep[hot spots and jets; see][]{lp95} which may also be sources of  additional optical continuum emission, since \cite{ar10} showed that the radio emission in \object{3C 390.3} is correlated with the optical continuum.

(iii) \textit{NLSy 1  \object{Ark 564}:}  The amplitude of variability of \object{Ark 564} is small, and consequently the corresponding dynamical range (EW versus L5100) is relatively small, which may flatten the Beff slope. However the intrinsic Beff in this AGN is clearly detected. We note that the variability in high-energy light curves shows remarkably high amplitude \citep[see, e.g.,][]{ka13}. Some NLSy 1 are peculiar AGNs, the variability in the optical spectra shows many flare-like features \citep[see, e.g.,][]{sh12}, and they may be $\gamma$-ray emitters, indicating the presence of relativistic jets or outflows like those  in blazars \citep[see][]{ya15}. \object{Ark 564} shows a weak correlation between $\log$(EW) and continuum (Table \ref{res}) and there is a difference in the slope and correlation coefficient for H$\alpha$ and H$\beta$, which is expected since our long-term monitoring showed a weak correlation between H$\alpha$ and H$\beta$ fluxes \citep[see][]{sh12}.

(iv) \textit{High-luminosity quasar  \object{E1821+643}:} It is clear that \object{E1821+643} exhibits significant intrinsic Beff as seen in other objects in our sample. The quasar \object{E1821+643} also shows change in the slope of the intrinsics Beff.

\subsection{ H$\alpha$/H$\beta$ flux ratio versus continuum}

\begin{figure}
        \centering
        \includegraphics[width=\columnwidth]{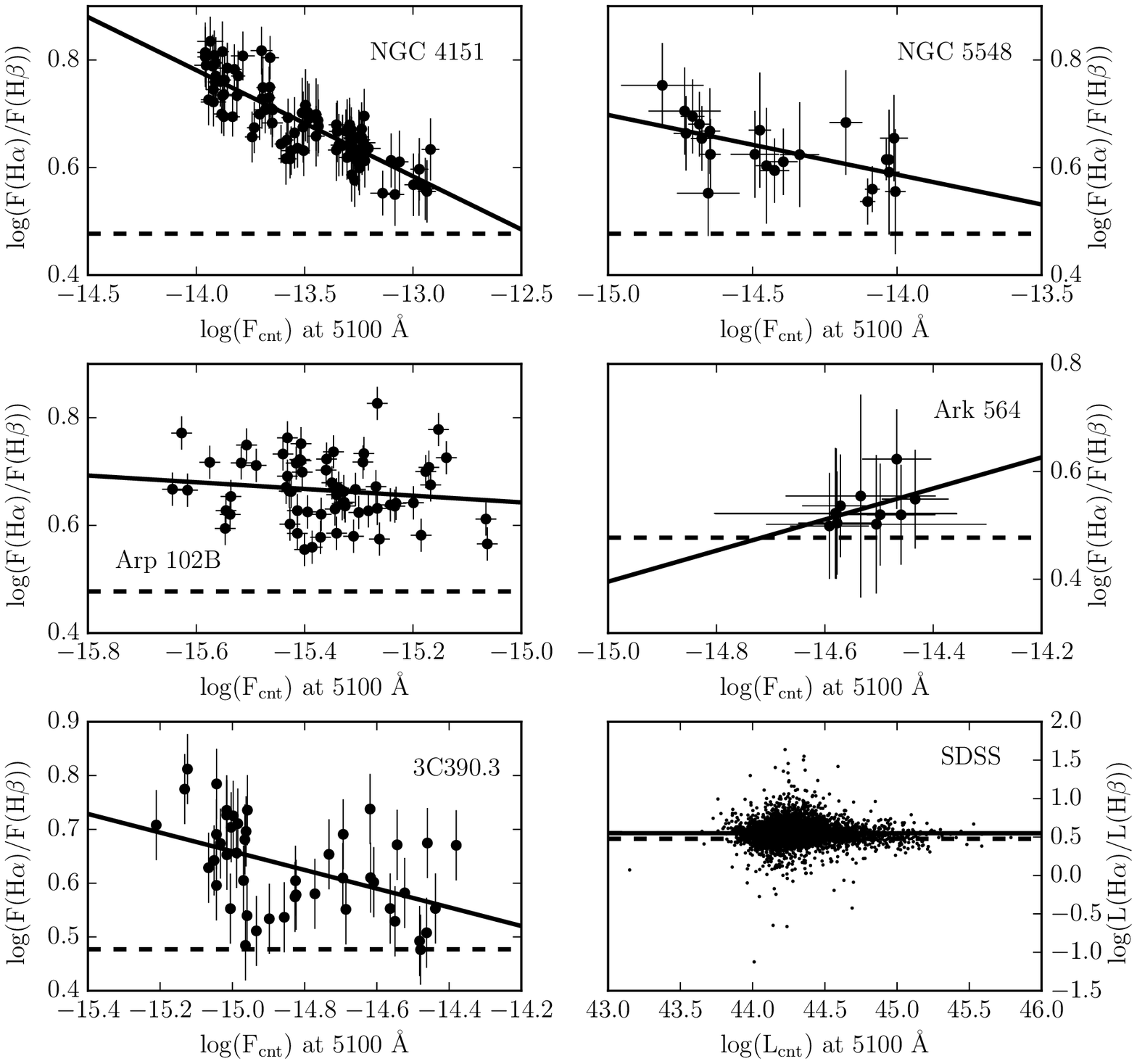}   
        \caption{H$\alpha$/H$\beta$ flux ratio vs. the continuum 
flux at 5100 \AA\, in logarithm scale, for the five AGNs in our sample for which both H$\alpha$ and H$\beta$ fluxes are available (name denoted on each plot), and for a sample of 4800 SDSS AGNs from \cite{shen11} (bottom right panel). The solid line represents the best fit of the observed data and the dashed line represents 
the typical ratio H$\alpha$/H$\beta\approx 3$ expected in the case of pure photoionization.}
        \label{ratio} 
\end{figure}
In addition to the intrinsic Beff, we explore the H$\alpha$/H$\beta$ flux ratio as a function of the continuum flux. Assuming a pure photoionization model, we can expect that the H$\alpha$/H$\beta$ broad line ratio  has
an average value of $\sim 3$ \citep{of06}, as observed in a number of AGNs \citep{do08}. On the other hand,
the ratio of  H$\alpha$/H$\beta$ can give an indication of some physical processes in the BLR \citep[see][]{po03,po08,il12} that might be connected with the intrinsic Beff.

The H$\alpha$/H$\beta$ line flux ratios versus the continuum 
flux at 5100 \AA\ for all objects, except \object{E1821+643} for which the H$\alpha$ line is not available, are shown in Fig. \ref{ratio}. We also plot in Fig. \ref{ratio} (bottom  right panel) the H$\alpha$/H$\beta$ versus  continuum luminosity at 5100 \AA\ for a subsample of 4800 objects (for which H$\alpha$, H$\beta$, and continuum at 5100 \AA\  luminosities were available) from the SDSS database studied by \cite{shen11}.  It can be seen in Fig. \ref{ratio} that the behavior of the $\log$(H$\alpha$/H$\beta$) versus continuum is different for different  AGNs in our sample, but it is always above 3 (dashed line in all panels).
It seems that the H$\alpha$/H$\beta$ anti-correlates with the continuum emission (solid line in the top left panel) only in \object{NGC 4151}. For the rest of the AGNs  there is no significant (anti-)correlation with the continuum, and the H$\alpha$/H$\beta$ ratio has different responses to the continuum: i) in the case of \object{NGC 5548} there is a similar trend to that in the case of \object{NGC 4151}, but with higher scatter; ii) in  \object{Arp 102B} and \object{Ark 564} the ratio is more or less constant; and iii) in \object{3C 390.3} the change is irregular, which may be due to two different variability phases of the broad line profiles, one caused by the central source outburst and second caused by the change in the disk structure \citep[see][]{jo10,po11}. Finally, the averaged H$\alpha$/H$\beta$ ratio (solid line in the bottom right panel) in the sample of 4800 AGNs is almost 3.

 The most important aspect to be noted is that although the intrinsic Beff is present in 
all objects, the H$\alpha$/H$\beta$ flux ratio varies differently with the continuum for different 
AGNs in our sample. In the case of \object{NGC 4151} and \object{NGC 5548}, this ratio  tends to  3 for the highest flux states, and for the case of \object{Ark 564} is almost constant and close to 3, which is expected as the photoionization is dominant. However, in the case of \object{Arp 102B} and \object{3C 390.3} (AGNs with double-peaked lines) this ratio as a function of the optical continuum seems to be more complex.

Finally, we note that the H$\alpha$/H$\beta$ line flux ratio in AGNs may be complicated and in general is not fully understood;  it may be also affected by the reddening and optical depth effect. However, in the global picture (see bottom right panel in Fig. 4), no trend similar to the trend seen in \object{NGC 4151} is present between the H$\alpha$/H$\beta$ ratio and the luminosity.

\subsection{Global Beff}

\begin{figure}  
        \centering
        \includegraphics[width=\columnwidth]{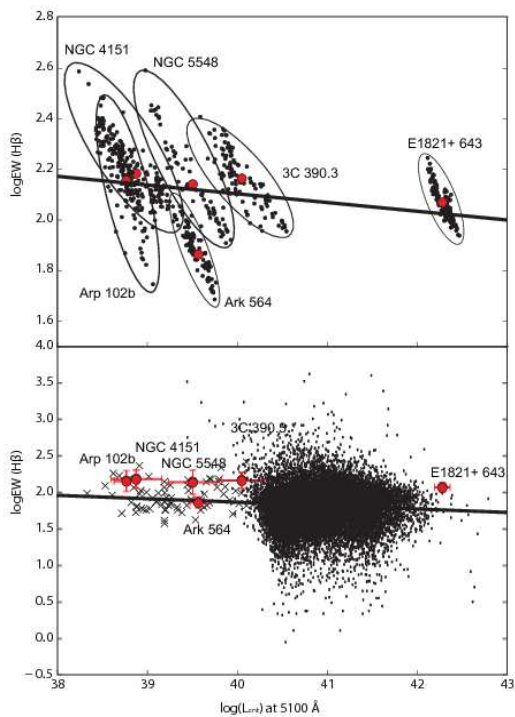}        
        \caption{Global Beff for H$\beta$. Top: For our samples of six AGNs, we plot the H$\beta$ line EW and continuum luminosity at 5100 \AA \, (dots surrounded with contours). The corresponding averaged line EW and continuum luminosity is denoted with a circle for each object. The best-fitting of all the measurements for the six objects is shown with the solid line. Bottom: Averaged values of the six studied objects (circles), together with the sample of 21416 AGNs measured by \citet{shen11} (dots) and 90 low-luminosity AGNs from \cite{la07} (crosses). The best-fitting of all data is shown with the solid line.}
        \label{gl} 
\end{figure}

It is interesting to consider the intrinsic Beff in the context of the global one, since  
the broad Balmer lines do not show global Beff \citep[see, e.g.,][]{kov10,pk11}.

To compare the intrinsic and global Beff, we plot the EW of the H$\beta$ line versus the continuum 
luminosity at 5100 \mbox{\AA}  for all measurements of the six considered AGNs 
(dots surrounded by contours in the top panel of Fig. \ref{gl}). 
The circles in both panels in Fig. \ref{gl} represent the average value of EW and continuum. 
From this plot, it seems that the global Beff can hardly hold in 
the broad component of the H$\beta$ emission line when all six objects are considered 
(full line in Fig. \ref{gl}).

This situation is even more noticeable when these data are compared with the results obtained for a large sample of 21416 sources from \cite{shen11}, and with 90 low-luminosity sources from \cite{la07} (Fig. \ref{gl}, bottom
panel). As Fig. \ref{gl} clearly illustrates, even though the EWs measured in this large sample are fully consistent with those estimated from the variability monitoring campaign,
no evidence of a significant global Beff in the broad component of H$\beta$ is detectable  (slope $\beta=-0.0467$). 

 We can conclude that according to  the results
for the six monitored AGNs and for the much larger sample of single-epoch observations (Fig. \ref{gl}), we 
could hardly detect significant global Beff in the broad component of $\rm H\beta$. 
This agrees with previous findings, which concluded that no global Beff was 
present in the broad Balmer lines \citep{di02,kov10}.

\section{Physical interpretation of the intrinsic Beff}

 Before discussing the possible physical  explanations of the intrinsic Beff, we summarize several facts noted above:

(i) There is a significant intrinsic Beff in all six AGNs, which makes only little difference between the different types of AGNs. The broad line profiles of the considered AGNs are quite different, and the structure of the BLRs seems to be different. Consequently, the geometry of the BLR probably does not have a significant impact on the presence of the intrinsic Beff;

(ii) In an individual AGN, there are different Beff slopes in different periods defined on the basis of the continuum flux (going from the high to the low activity phase, or vice versa). This may be connected with the nature of the ionizing continuum source and consequently with the physics of the BLR;

(iii) There is a different response of the H$\alpha$/H$\beta$ ratio for different objects. This may
be caused by the different mechanism of the optical continuum emission, but also by the change in the structure of the line-emitting disk-like region \citep[e.g., in 3C390.3; see][]{po11} 
or spatially extended BLR \citep[see][]{gk14};

(iv) The intrinsic Beff is not connected with the global Beff, i.e., even if the global Beff is not present in the broad Balmer lines, the intrinsic Beff is significant.

 Moreover, Seyfert 1 and NLSy1 AGNs show different spectral characteristics caused by different inclinations, accretion rates, and other physical properties \citep[see, e.g.,][etc.]{sul00, sul09, zh06, sh14, bi17}, which should probably result in a different intrinsic Beff. But this is not the case when comparing the intrinsic Beff of \object{Ark 564} (NLSy1) and other Type 1 AGNs in our sample.
Therefore, we can conclude that the intrinsic Beff is probably not connected with the orientation or the accretion rate. However, we should note that we have only one NLSy1 in our sample, and the influence of the accretion rate should be checked on a larger sample of NLSy1 AGNs.

We considered AGNs with quite different broad line characteristics.
First, in two Seyfert 1 galaxies \object{NGC 4151} and \object{NGC 5548} in addition to the line and continuum variability, the shape of line profiles has also varied significantly, especially in the case of \object{NGC 4151} \citep{sh10a}. 
This variability could be associated with the contribution of outflows \citep{il10} or even with the 
SMBBH \citep{bo12,li16}. Then, \object{3C 390.3} seems to be  representative of the disk-like BLR \citep[][]{po11}, while this is not clear in the case of \object{Arp 102B} \citep{po14}, even though they both have double-peaked broad Balmer lines. Additionally, we have the extremely luminous quasar \object{E1821+643} with strongly redshifted broad lines, which is regarded as a SMBBH candidate \citep[see][and references therein]{sh16}.  Finally, \object{Ark 564} is an NLSy1 galaxy whose BLR might have a different geometry.  Hence, a question arises: What do all of these AGNs that can produce the intrinsic Beff have in common?

 To find an answer to this question, we should explore the physical causes connected with the central continuum source and physical properties in the BLR. We consider the photoionization, recombination, and collision effects, taking into account that for higher ionization parameters, the photoionization followed by recombination is more important in the BLR. However at higher temperatures or in the case of low-ionization parameters  we can expect that the collisional excitation becomes important \citep{of06,po08,il12}. Apart from these two effects, the radiative-transfer effects in Balmer lines can affect the broad Balmer line ratio. However, we can assume that  photoionization is the dominant process in the BLR, and so to explain the observed intrinsic Beff we start with  photoionization.

\subsection{Pure photoionization model}
\label{cl}
 
To explore the physical background of the intrinsic Beff, we first generated several grids of the BLR models using the photoionization code Cloudy \citep[version 13.03,  described in][]{fe13}. 

\begin{figure}
        \centering
        \includegraphics[width=\columnwidth]{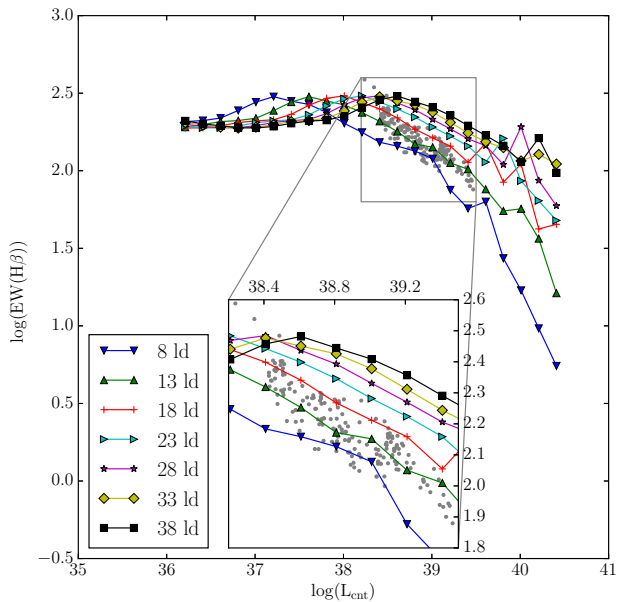}
        \caption{Correlation of the $\rm EW \,H\beta$ with the incident luminosity of the continuum
        at 5100 \AA\, for different $\rm R_{BLR}$ (denoted in the bottom left), obtained from the Cloudy  simulations. Circles represent observed intrinsic Beff for the $\rm H\beta$ line of \object{NGC 4151} with continuum fluxes corrected for the host-galaxy contribution (zoomed-in for better visibility).}     
        \label{fig:tot}
\end{figure}

We modeled  a simple single BLR cloud in an open geometry, as plane parallel slabs shed by the continuum 
of an AGN, defined through the bolometric luminosity ($\rm L_{bol}$), the radius of the BLR ($\rm R_{BLR}$), and the hydrogen density ($\rm n_H$).  We used the column density value of  $10^{23}\,\rm cm^{-2}$ for the stopping criteria. The hydrogen density was fixed to $\rm n_H= 10^{11}\ cm^{-3}$ and the shape of the ionizing continuum was chosen as the table of the characteristic continuum for an AGN \citep{fe98}.  To produce one grid of models we varied $\rm \log L_{bol}$ within the range [41-45.2] $\rm erg\,s^{-1}$ with an intermediate step of 0.2 dex.

We attempted to explain the intrinsic Beff in \object{NGC 4151} that had a relatively compact BLR \citep[see][]{pc88,cl90,ma91,be06,sh08}; therefore, we produced several grids of models by varying the $\rm R_{BLR}$ in the range of 8-38 light days (l.d.), with a step of 5 l.d.
The EW was calculated in a similar way using 
Eq. \ref{eq:ew}, but in this case we used luminosities instead of fluxes.
In Fig. \ref{fig:tot} we plot the EW(H$\beta$) versus the incident continuum luminosity at 5100 \AA\ from the output of the Cloudy code calculation. We explore their correlation in order to assess whether an intrinsic Beff is predicted by a simple photoionization theory, and to identify
the origin of different Beff slopes of a single source. We note that detailed studies of variability of broad lines under photoionization processes are given in \citet{ko04} and \citet{gk14},
where different effects are considered. Here, we used a simple model in order to reproduce the effects observed in the case of \object{NGC 4151}.

\subsubsection{NGC 4151:  a study case}

We tried to reproduce the observed Beff of \object{NGC 4151}, as it is shown in Fig. \ref{baldwin} panel 7, through our set of photoionization models (see Fig. \ref{fig:tot}). Figure \ref{fig:tot} shows that a strong intrinsic Beff (with a slope of -0.331 and -0.205) is present for clouds located close to the source, i.e., for models obtained with $R_{BLR} = 8$ l.d. and $R_{BLR} = 13$ l.d. respectively. At larger distances  (i.e., larger BLR radius), for the low-luminosity states, the EW of H$\beta$ is nearly constant, but for higher luminosity states there is a significant drop in  EW and an intrinsic Beff occurs. Taking into account that the modeled clouds have fixed column density (which determines the cloud size), the emitted line intensity (which responds to an increasing continuum flux) will grow until there is no neutral gas in the cloud that can be further ionized. This corresponds to the nearly constant EW that is observed in the low-luminosity state (Fig. \ref{fig:tot}). When the flux is high enough to completely ionize the cloud, the line intensity can no longer grow with the continuum and a sharp drop in the EW is present. Since the flux grows linearly with luminosity, but decreases with the square of the cloud distance, the drop will appear at higher luminosities in more distant clouds. On the other hand, the cloud will see a large flux at low luminosity when it is closer to the source (so EW will be larger at small radii), but it will also saturate sooner; i.e., when the luminosity increases, the EW of a closer cloud decreases before the EW of a farther one.
Figure \ref{fig:tot} implies that one of the reasons for the variability of the intrinsic Beff in \object{NGC 4151} (illustrated in Fig. \ref{baldwin} panel 8) could be a change of the line emission site in the BLR.  A comparison of the models with observed continuum luminosities (plotted as circles in Fig. \ref{fig:tot}), suggests that agreement can only be achieved over a limited range of BLR radii.

Based on the simulations, the possible observed variability in the intrinsic Beff could be explained with the changes in the size of the emitting region and with different gas layers contributing to the various components of the line profile, depending on the strength of the incident continuum. However, the BLR of \object{NGC 4151} seems to be very complex \citep{il08,sh08,sh10a,bo12} and we have to consider that photoionization might not be the only mechanism of ionization \citep{sh08}. One problem of particular interest in this AGN is the determination of its BLR size. A number of reverberation studies have shown inconsistent results for the BLR dimension, finding that the BLR may extend from less than 2 l.d. \citep{sh08} to 4--10 l.d. \citep{pc88,cl90,ma91,be06}, but all of these studies agreed that the BLR was very compact. Our models show that the intrinsic Beff strength could be explained with photoionization, but only if the BLR size is between 8 and 13 l.d. (see Fig. \ref{fig:tot}), in agreement with the previous results of the compact BLR. We note here that \citet{sh08} showed, using the grid of Cloudy models given by \citet{ko97},  that the observed H$\beta$ line fluxes could be explained only by a larger BLR ($\sim$30 l.d., which  can emit  an intensive line flux that is far larger than the reverberation estimates) or by an additional ionizing continuum responsible for the line production, which comes from either outflows or jets.

Although the observed slope of the intrinsic Beff for the H$\beta$ line in \object{NGC 4151} (p1+p2: -0.392) is close to the values predicted by models with $R_{BLR}$ = 8 and 13 l.d. (-0.331 and -0.205, respectively; see Fig. \ref{fig:tot}), the models do not fit  the observations exactly, but these simple models give us an idea of the possible reasons for changes in the intrinsic Beff slope. The idea is that the emission line site in the BLR  moves on   radii that are farther away  with an increase in the continuum luminosity; i.e., a ``breathing effect” exists in the BLR of \object{NGC 4151}. A similar effect was detected in the BLR of \object{NGC 5548} \citep{ch06}.

\subsubsection{Photoionization and the intrinsic Beff: possible scenario and limits}

The proposed simple photoionization models provide a possible scenario which could explain the intrinsic Beff: as the ionization continuum increases, the broad line emission becomes saturated and the inner layers of the BLR become optically thin for the continuum radiation. As a consequence, the site of line formation moves to more distant gas layers, which react to the changes in the incident photon flux with a different response. 
\begin{figure}
        \centering
        \includegraphics[width=\columnwidth]{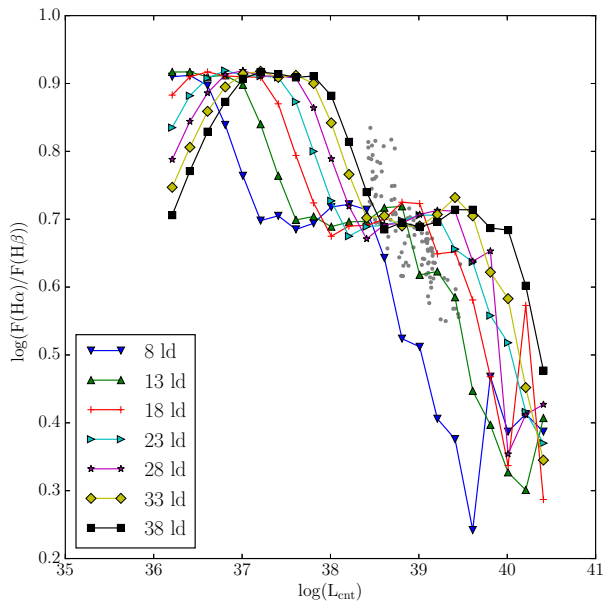}
        \caption{H$\alpha$/H$\beta$  ratio vs. the continuum at 5100 \AA\ for different $\rm R_{BLR}$ sizes (denoted in the bottom left) compared with the  observed ratio for \object{NGC 4151} (circles).}       
        \label{ratio1}
\end{figure}

However, the simple photoionization models have several limitations that we  discuss here. The first  is the ratio of Ly$\alpha$ to H$\beta$ obtained from our simulations that is above 30, in contradiction with the measured values around 20-25 \citep[][]{pe79,ne95}. A second point is that we cannot reproduce the H$\alpha$/H$\beta$ ratio. As  can be seen in Fig. \ref{ratio1}, the modeled ratios do not fit the observed H$\alpha$/H$\beta$ ratios (circles), especially in the low-activity state of \object{NGC 4151}.  In contrast to the usual nebular case, the H$\alpha$/H$\beta$  ratio can span  a wide range of values in a BLR environment; for example, several Type 1 AGNs exhibit ratios H$\alpha$/H$\beta > 3$. In BLR clouds there are several factors that affect this ratio: superthermal electrons due to the high-energy AGN radiation field, trapped Ly$\alpha$ photons that overpopulate hydrogen in its level 2, enhanced recombination due to high density, and a possible contribution from an intrinsic extinction. All these effects combine in such a way that we expect higher H$\alpha$/H$\beta$ ratios with respect to a Case B situation whenever the cloud has a significant fraction of neutral gas. Most of the models actually appear to predict a very high H$\alpha$/H$\beta$ ratio in the low-flux regime, perhaps indicating that the high-order Balmer lines are destroyed in neutral gas, favoring H$\alpha$ much like what happens in the Lyman series at lower densities. When the flux grows (because of the luminosity increase, but the effect naturally occurs at higher luminosities for more distant clouds), the destruction of high-order Balmer photons becomes ineffective and the ratio reaches a quasi-constant value (as it can be seen in Fig. \ref{ratio1}). In our models, this occurs approximately in the same luminosity range where the EW achieves its maximum, meaning that this is the range of optimal conditions for the emission of H$\beta$. If, however, the continuum grows at even higher fluxes, the H$\alpha$/H$\beta$  ratio experiences a quick drop. This effect can occur because clouds in such conditions rapidly become inefficient line emitters, leading to emission line ratios that tend to even out.
\begin{figure}
        \centering
        \includegraphics[width=\columnwidth]{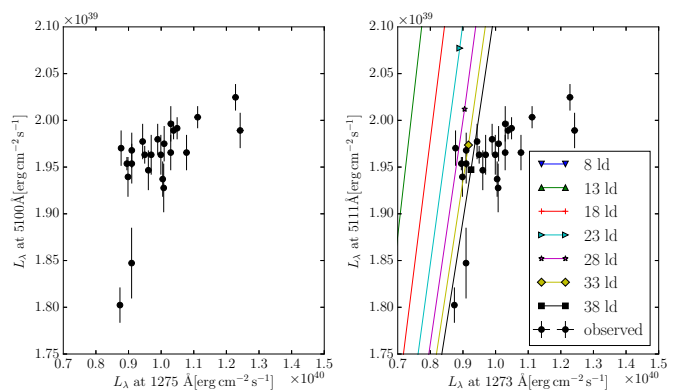}
        \caption {Left: Observed continuum flux at 1275 \mbox{\AA} vs. the flux at 5100 \mbox{\AA} for \object{NGC 4151} obtained during short-time period (see text for details). Right: Modeled continuum flux at 1275 \AA\ vs. the flux at 5100 \AA\, (solid lines for different $\rm R_{BLR}$ sizes, denoted in the bottom left) compared to the observed values (full circles).}
        \label{uvopt}
\end{figure}

A further concern for simple photoionization calculations is the  nontrivial relationship existing between the optical continuum and the intensity of the ionizing radiation field. We tested the relation between the ionizing flux in the UV and optical spectral bands in the case of \object{NGC 4151}, using the data from short-timescale multiwavelength observations carried out by \cite{cren96uv} (UV) and \cite{kaspi96opt} (optical) in December 1993. In Fig. \ref{uvopt}, we show the relation between the continua at 1275 \AA \, and 5100 \AA. We calculated the Pearson correlation coefficient $r = 0.577$ with $P = 0.003$ and we performed the best linear fit obtaining a slope of 0.028. Additionally, we compared the UV versus optical continuum with those from photoionization models. As  can be seen in Fig. \ref{uvopt}, although there is some degree of correlation between the two continua, strong variations in the UV are only followed by weak changes in the optical domain. Furthermore, in the low-luminosity stage there appears to be a large break, which would be unexpected in the case of pure photoionization, but could be explained by an optical continuum component that does not relate with UV. This becomes apparent when the AGN is in its low-activity stage and the ionizing continuum is not dominating the overall spectrum.

It seems that the photoionization model, which changes  the BLR size, can explain the intrinsic Beff, but there are several observational  facts (the H$\alpha$/H$\beta$ ratio and UV versus optical continuum variability) that cannot be explained by this model with the same parameters.  These  differences might be accounted for by the presence of effects that were not considered in the photoionization calculations.

\subsection{Additional continuum emission}
First, we recall some results obtained from variability investigations of the AGNs in our sample.
We start with the conclusions in \citet{sh08}, where the photoionization model could not explain the correlation between the line and continuum flux in \object{NGC 4151} since  a saturation in the broad line flux was observed in the high continuum state \citep[see Fig. 10 in][and discussion in the paper]{sh08}. It seems that an extra ionizing continuum source is present  in \object{NGC 4151}.  We can also see in Fig. \ref{uvopt} that there is no significant response of the optical to the UV continuum during the short period. 
In addition, in the case of 3C390.3 it was discovered that the variability in the optical continuum had a strong correlation with that in  the radio continuum \citep[see][]{ar10}. More recently \cite{gk15} considered the EW variation of \object{NGC 5548}, and in order to explain the EW variability they invoked a smaller value of the incident ionizing photon flux that was different from the observed UV continuum flux and typical models of the continuum SED. 

Taking into account all the above, we should consider a possibility that the intrinsic Beff may be caused by  a portion of additional continuum emission that is not connected with the ionization continuum. The total optical continuum can be considered as the sum of the optical continuum that follows the SED of the UV ionizing continuum, and a fraction of the additional optical (nonionizing) continuum that can be produced by different processes.
The luminosity of broad lines $L_{\rm line}$ is a function of the ionizing continuum \citep[see, e.g.,][]{ko04} as
 \begin{equation}
 L_{\rm line}=A\cdot L_{\rm ion}^\alpha,
 \end{equation}
  where $\alpha$ represents an emission-line responsivity, giving
 \begin{equation}
\log EW \sim (\alpha-1) \log L_{\rm ion}.
 \end{equation}

The expected intrinsic Beff slope $\beta$ should not be too high; i.e., assuming photoionization it follows that  $R_{\rm BLR}\sim L_{}^{0.5}$ \citep{be06b}, and using a rough estimate that $L_{\rm line} \sim R_{\rm BLR}^3$ \citep{of06}, we obtain $\alpha\sim 1.5$, giving $\beta\sim0.5$. In contrast, we infer that in some cases $\beta$ is around -1 (see Table \ref{res}), indicating very low values of $\alpha$. This may imply that  the reprocessing of the ionizing continuum by the line emission is very low and part of this continuum is transformed into the optical one, which is observed as the additional continuum in the optical band. This is also in agreement with the results obtained for \object{NGC 5548} broad line variability by \cite{gk15}.

A part of the additional optical continuum may be emitted partially from the BLR itself in free-bound and free-free transitions since, with the increase in the ionization continuum, we can expect an increase in the rate of ionization, i.e., the number density of electrons and ions. However, the additional continuum could also come from jet or outflow, as  in the case of \object{3C 390.3} \citep[see][]{ar10}.  We note here that the ``additional continuum'' problem has to be considered in more detail, taking into account the X-ray, UV, and optical spectra of several AGNs, which is beyond the scope of this paper.

\section{Conclusions}
\label{sec:con}

The intrinsic Baldwin effect in the broad H$\beta$ and H$\alpha$ lines of a sample of six Type 1 AGNs was studied. In our sample, we considered two Seyfert 1 galaxies (\object{NGC 5548}, 
\object{NGC 4151}), two AGNs with very broad double-peaked lines (\object{3C 390.3} and \object{Arp 102B}), one NLSy 1 galaxy
(\object{Ark 564}), and one high-luminosity quasar (\object{E1821+643}). We used data obtained from a single long-term monitoring campaign \citep{sh04,sh08,sh10b,sh12,sh13,sh16}, which made our sample uniform and consistent. We compared observed data with the results obtained with the numerical code Cloudy, and also discussed the global Beff of these six AGNs,  combined with the sample of $\sim$21500 objects from the SDSS database. We can outline the following conclusions:
\begin{enumerate}
        \item  The intrinsic Beff is present in broad Balmer lines of six Type 1 AGNs 
studied in this paper. The shift and the change in the slope of the intrinsic Beff is observed in all the considered AGNs. 
Taking into account that the AGNs in the studied sample have different line shapes that indicate different BLR geometries, the intrinsic Beff is probably not caused by the geometry of the BLR. However, this should be tested on larger samples of different AGNs.
        \item The lack of a significant global Beff for the broad Balmer lines is confirmed, and there is        no connection between the global and intrinsic Beff. 
        \item  Using the Cloudy code to build a simple model of line emitting clouds in the BLR of \object{NGC 4151}, we found that the photoionization models were able to reproduce the intrinsic Beff, which also suggests that the 
variation in the Beff slope in \object{NGC 4151} might be due to a change of the site of line formation in the BLR.

However,  pure photoionization cannot explain the H$\alpha$/H$\beta$ ratio and the   Ly$\alpha$/H$\beta$ ratio in this object. Consequently, the additional optical (nonionizing) continuum that may be emitted from the BLR or from outflows can be an alternative, but the natures of these two possibilities have to be investigated in greater detail.
\end{enumerate}

Although our considerations of additional continuum components affecting the Beff are mainly based on indirect evidence, we recall that deviations from a linear line response to the underlying continuum have been observed in \object{NGC 4151}  in both its highest and lowest activity. In order to constrain the contribution of nonionizing continuum components, it is  important to study AGNs  in their maximum and their minimum activity when deviations of broad line responses are expected to be more evident.

\section*{Acknowledgements}
We are grateful to the anonymous referee for carefully reading our
paper and giving comments and suggestions that greatly improved our paper.
This work is a part of the project (176001) ``Astrophysical Spectroscopy of Extragalactic Objects'',
supported by the Ministry of Education, Science and Technological Development of Serbia and the project
``Investigation of supermassive binary black holes in the optical and X-ray spectra'' supported by the  
Ministry of Science and Technology of R. Srpska. Also, this work was supported by INTAS (grant N96-0328) and RFBR (grants N97-02-17625
N00-02-16272, N03-02-17123, 06-02-16843, N09-02-01136,12-02-00857a, 12-02-01237a, N15-02-02101). 
The work was partly  supported by the Erasmus Mundus Master Program, AstroMundus.

\bibliography{baldwin}

\end{document}